\documentclass{article}
\newcommand{\eff}{\mbox{\footnotesize eff}}
\newcommand{\mf}[1]{\mbox{\footnotesize #1}}
\renewcommand{\(}{\left (}
\renewcommand{\)}{\right )}
\newcommand{\beq}{\begin{equation}}
\newcommand{\eeq}{\end{equation}}
\newcommand{\beqa}{\begin{eqnarray}}
\newcommand{\eeqa}{\end{eqnarray}}
\newcommand{\sss}{{\bf s}}
\newcommand{\vv}{{\bf v}}
\newcommand{\PP}{{\bf P}}
\newcommand{\LL}{{\bf L}}
\newcommand{\uu}{{\bf u}}
\newcommand{\si}{\sss_i}

\newcommand{\vi}{\vv_i}
\newcommand{\ui}{\uu_i}

\newcommand{\sij}{s_{ij}}
\newcommand{\vij}{v_{ij}}
\newcommand{\uij}{u_{ij}}
\newcommand{\Pij}{P_{ij}}
\input{psfig}

\begin{document}
\begin{titlepage}
\begin{flushright}
LU TP 97-10 \\
\today \\
\end{flushright}
\vspace{0.8in}
\LARGE
\begin{center}
{\bf Airline Crew Scheduling Using\\ Potts Mean Field Techniques}\\
\vspace{0.3in}
\large
Martin Lagerholm\footnote{martin@thep.lu.se},
Carsten Peterson\footnote{carsten@thep.lu.se} and
Bo S\"oderberg\footnote{bs@thep.lu.se}\\
\vspace{0.1in}
Complex Systems Group, Department of Theoretical Physics\\
University of Lund, S\"olvegatan 14A, S-223 62 Lund, Sweden\\

\vspace{0.3in}

Submitted to {\it Operations Research}

\end{center}
\vspace{0.2in}
\normalsize

Abstract:

A novel method is presented and explored within the framework of Potts
neural networks for solving optimization problems with a non-trivial
topology, with the airline crew scheduling problem as a target
application. The key ingredient to handle the topological
complications is a propagator defined in terms of Potts neurons. The
approach is tested on artificial problems generated with two
real-world problems as templates.  The results are compared against
the properties of the corresponding unrestricted problems. The latter
are subject to a detailed analysis in a companion paper \cite{loc}.
Very good results are obtained for a variety of problem sizes. The
computer time demand for the approach only grows like $\mbox{(number
of flights)}^3$. A realistic problem typically is solved within
minutes, partly due to a prior reduction of the problem size, based on
an analysis of the local arrival/departure structure at the single
airports.

To facilitate the reading for audiences not familiar with Potts neurons and
mean field techniques, a brief review is given of recent advances
in their application to resource allocation problems.

\end{titlepage}

\newpage

\section{Introduction}

Artificial Neural Networks ({\bf ANN}) have over the last decade emerged as
powerful tools for ``intelligent'' computing. Most attention has been
paid to feed-forward architectures for pattern recognition and
prediction problems. Conceptually, these approaches tie nicely into
existing statistical and interpolation/extrapolation schemes.  The
application of feedback ANN methods to combinatorial optimization
problems \cite{tank,pet2,dur,pet8} also looks very promising. In
contrast to most search and heuristics methods, the ANN-based approach
to optimization does not fully or partly explore the space of
possible configurations; rather, the ANN ``feels'' its way through a
continuous space of fuzzy configurations towards a good final
solution. The intermediate fuzzy configurations have a natural
probabilistic interpretation.

Typically, two basic steps are involved when using ANN to find good
solutions to combinatorial optimization problems \cite{lenstra}: (1)
map the problem onto a neural network (spin) system with a
problem-specific energy function, and (2) minimize the energy by means
of a deterministic process based on the iteration of mean field (MF)
equations.

Initially, most applications concerned fairly artificial problems like
the traveling salesman problem, various graph partition problems
\cite{tank,pet2} and knapsack problems \cite{ohl,ohl1}. In
refs. \cite{gis1,gis2}, a more realistic problem (high school
scheduling) was addressed. In all these applications, topological
complication was not an issue, and could be dealt with in a
straightforward way using ``standard'' ANN energy functions similar to
those encountered in spin physics.

Recently, a formalism has been developed within the feedback ANN
paradigm to handle applications with more complicated topologies, like
airline crew scheduling and telecommunication routing problems
\cite{lager,hakk}. This paper deals with airline crew scheduling using
the techniques briefly reported in ref. \cite{lager}.

In airline crew scheduling, a given flight schedule is to be covered
by a set of crew {\it rotations}, each consisting in a connected
sequence of flights ({\it legs}), that begins and ends at a
distinguished airport, the {\it home base} ({\bf HB}). The total crew
time is to be minimized, subject to a number of restrictions on the
rotations.

A commonly used approach to this problem proceeds in two steps.  (1)
First a large pool of feasible crew rotations that conform with the
restrictions is generated (this is often referred to as column or
matrix generation). (2) With such a set as a starting point, the
problem is then reformulated as finding the best subset of rotations
such that each flight is covered precisely once. This transforms the
problem into a {\it set partitioning} problem (see e.g. \cite{hoff}
and references therein). Solutions to this ``standard'' problem are
then found by approximate methods based on e.g. linear programming;
more recently an exact branch-and-cut method has been used
\cite{hoff}.  Even for moderate problem sizes, feasible rotations
exist in astronomical numbers, and the pool has to be incomplete; this
approach is therefore non-exhaustive.

Feedback ANN methods could be used to attack the resulting set
partitioning problem. In fact, ANN methods have been successfully
applied to the similar knapsack and set covering problems
\cite{ohl,ohl1,ohl2}. We will, however, follow a completely different
pathway in approaching the airline crew scheduling problem: First, the
full solution space is narrowed down using a reduction technique that
removes a large part of the sub-optimal solutions.  Then, an MF
annealing approach based on a Potts neuron encoding is applied. A key
feature here is the use of a recently developed propagator formalism
\cite{lager} for handling topology, leg-counting, etc.

The method is explored on a set of synthetic problems, which are
generated to resemble two real-world problems representing long and
medium distance services. The algorithm performs well with respect to
solution quality, with a computational requirement that at worst grows
like $N_f^3$, where $N_f$ is the number of flights.

The reduction technique employed, and the evaluation of the test
problem results, rely heavily upon exploiting the properties of the
solutions to the corresponding unconstrained problem, which decomposes
into a local problem at each airport, and is solvable in polynomial
time.  A fairly extensive analysis of these properties is given in a
companion article \cite{loc}.

This paper is organized as follows: In Section 2 we define the
problems under study, and Section 3 contains a discussion of the
properties of the unrestricted local problems. Our method for initial
reduction of the problem size is presented in Section 4. A generic
brief review of the art of mapping resource allocation problems onto
spin (neuron) systems, and a description of the MF annealing
procedure, can be found in Section 5, and in Section 6 the Potts MF
method for airline crew scheduling is presented. Section 7 contains
performance measurements on a set of test problems, and finally in
Section 8 we give a brief summary and outlook. Appendix A defines a
toy problem that is used throughout the paper for illustrating the
different techniques, while details on the Potts ANN algorithm and the
problem generator can be found in Appendices B and C respectively.

\section{Problem Definition}

In a realistic airline crew scheduling problem one wants to minimize
labour and other costs associated with a schedule of flights with
specified times and airports of departure and arrival, subject to a
number of safety and union constraints. Typically, a real-world flight
schedule has a basic period of one week.

The problem considered in this work is somewhat stripped. We limit
ourselves to minimizing the {\em total crew waiting-time}, subject to
the constraints:
\begin{itemize}
\item The crews must follow connected flight sequences -- {\em
	rotations} -- starting and ending at the home-base.
\item The number of flight legs in a rotation must not exceed a given
	upper bound.
\item The total duration (flight-time + waiting-time) of a rotation is
	similarly bounded.
\end{itemize}
We believe that these are the crucial and difficult constraints;
additional real-world constraints we have ignored do not constitute
further challenges from an algorithmic point of view.

Throughout this paper, we will use a small toy problem, depicted in
fig. \ref{fullrot}, to illustrate our approach. The underlying flight
data can be found in Appendix A.
\begin{figure} [htb]
\centering
\mbox{\psfig{figure=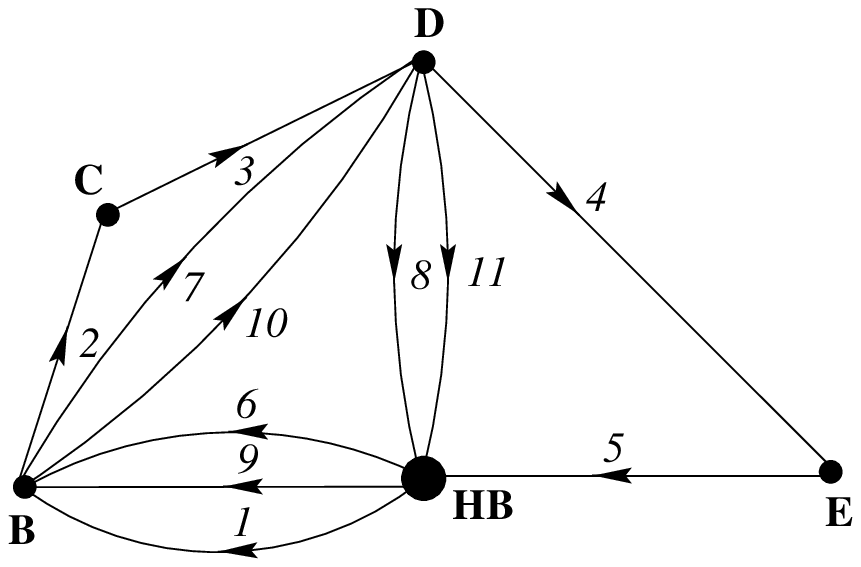 ,width=7.5cm}}
\caption{An illustration of the problem of table \protect\ref{fulldata}.}
\label{fullrot}
\end{figure}

Prior to developing our artificial neural network method, we will
describe a technique to simplify the problem, based on an analysis of
the local flight structure at each airport.

\section{Properties of the Unrestricted Solutions}

A solution to a given crew scheduling problem is specified by
providing, at each airport (except HB), a one-to-one mapping between
the arriving and departing flights. This implicitly defines the crew
rotations.

It is the global constraints that make the crew scheduling problem a
challenge.  In the absence of these, there will be no interaction
between the mappings at different airports; accordingly, the
waiting-times can be minimized {\em independently} at each
airport. This simplified problem will be referred to as the
corresponding {\em unrestricted problem}; it is solvable in polynomial
time. A detailed analysis of the statistical properties of such
problems is presented in ref. \cite{loc}. Here we briefly describe the
results from \cite{loc} needed for our preprocessing and analysis of
the results.

Summing the resulting minimal waiting-times over the airports defines
the minimal {\em unrestricted waiting-time}, denoted by
$T_{\mf{wait}}^{\mf{unr}}$.  This provides a lower bound to the
minimal waiting-time for the full problem.  Empirically, this bound is
almost always saturated, i.e. among the minimal solutions to the
unrestricted problem, a solution to the full problem can be
found. This can be understood as follows.

At a single airport, the waiting-time for a given mapping is obtained
by adding together the waiting-times for each arrival-departure pair 
($ij$), given by
\beq
	t_{ij}^{\mf{(w)}} =
	\( t_j^{\mf{(dep)}} -
	t_i^{\mf{(arr)}} \) \bmod \mbox{period}.
\eeq

Thus, the sum over pairs can only change by an integer number of
periods. At a large airport, the minimum often is highly degenerate:
For a random problem, the local ground-state degeneracy typically
scales as $\(N/2e\)^N$ for an airport with $N \gg 1$ departures per
period \cite{loc}. Consequently, the total number of minimal solutions
to a complete unrestricted problem, defined as the product of the
individual airport degeneracies, will be very large, and it is not
inconceivable that a solution satisfying the constraints can be found
among this set.

By insisting on ground-states, the state-space typically can be
reduced by a factor of two for each flight. Part of this reduction is
due to airports being split into smaller parts, which on the average
gives a factor of two for each airport. This will be exploited in the
next section, to reduce the size of a restricted problem. The
unrestricted ground-states will also be used when gauging the
performance of our Potts approach.

\section{Reduction of Problem Size}

By demanding a minimal waiting-time, the unrestricted local problem at
each airport (excluding the home-base) typically can be further split
up into independent {\em subproblems}, each containing a subset of the
arrivals and an equally large subset of the departures. Some of these
are trivial, forcing the crew of an arrival to continue to a
particular departure.

Similarly, by demanding a solution with $T^{\mf{unr}}_{\mf{wait}}$
also for the constrained global problem, this can be reduced as follows:
\begin{itemize}
\item {\em Airport fragmentation}: Divide each airport into {\em
	effective airports} corresponding to the unrestricted local
	subproblems.
\item {\em Flight clustering}: Join every forced sequence of flights
	into one effective {\em composite flight}, which will thus
	represent more than one leg and have a formal duration defined
	as the sum of the durations of its legs and the waiting-times
	between them.
\end{itemize}
Every problem will be preprocessed based on these two reduction
methods, which will be explained in more detail below.
In instances where no solution obeying the global constraints is found
within the reduced solution space, one can attempt to solve the
problem with no preprocessing. This was not necessary for any of the
probed problems.

\subsection{Airport Fragmentation}

\begin{figure} [htb]
\centering
\mbox{\psfig{figure=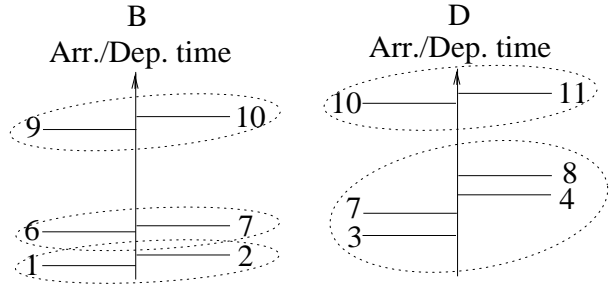,width=8cm}}
\caption{Arrival and departure times for airports B and D in the toy
example. The dotted ellipses mark the fragments into which the
airports can be divided.}
\label{times}
\end{figure}

Inspecting the local arrival and departure times reveals which
airports can be fragmented (for a full discussion, see
ref. \cite{loc}). In the toy example of fig.~\ref{fullrot}, airports
B and D can be split (see fig.~\ref{times}).
For airport B there is only one possibility for connecting the flights
without adding a period to the local waiting-time, yielding three
effective airports (B1, B2 and B3). Similarly, airport D can be
divided into two effective airports (D1 and D2). The structure that
results from this fragmentation is shown in fig. \ref{effrot}a.

\subsection{Flight Clustering}

\begin{figure}[t]
\centering
\mbox{  \psfig{figure=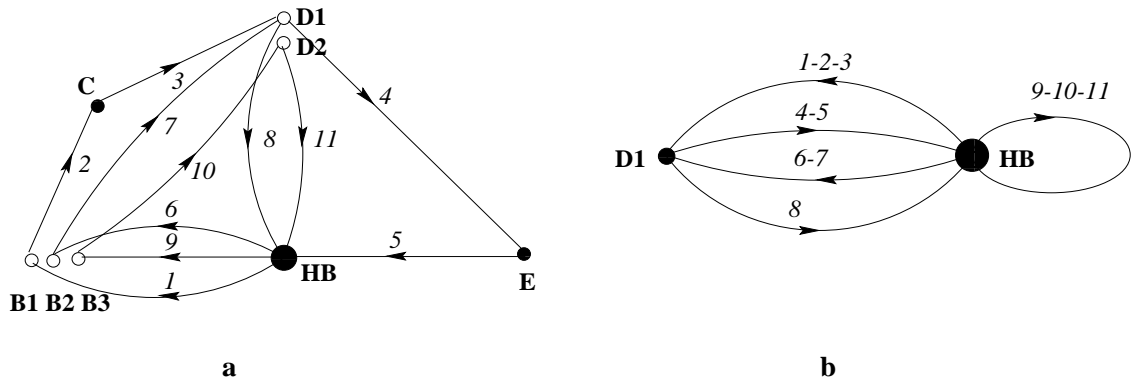,width=15cm}}
\caption{{\bf(a)} The effective airports resulting from airport
 fragmentation for the toy-problem, and {\bf (b)} the composite
 flights due to a subsequent flight clustering.}
\label{effrot}
\end{figure}

The airport fragmentation typically leads to several effective
airports having only one arrival flight and one departure
flight. Hence we can combine these into effective composite flights
(flight clustering), with a formal duration obtained by adding
together the flight duration times and the embedded waiting-times, and
an intrinsic leg-count given by the number of proper flights included.
The resulting structure for the toy problem is shown in
fig. \ref{effrot}b and table \ref{effdata}.

The {\em reduced problem} thus obtained differs from the original
problem only in an essential reduction of the sub-optimal part of the
solution space; the part with minimal waiting-time is unaffected. The
resulting information gain, taken as the natural logarithm of the
decrease in size of the solution space, empirically seems to scale
approximately like $2 \times$ (number of flights), and ranges from 100
to 2000 for the problems probed. This is considerably more than for a
completely random, unstructured problem, where the gain is expected to
scale like $\log 2 \times$ (number of airports) \cite{loc}.

\subsection{The Kernel Problem}

The reduced problem may in most cases be further separated into a set
of independent {\em subproblems}, connected only via the home-base;
these can be solved one by one. Some of the composite flights will
formally arrive at the same effective airport they started from. This
does not pose a problem; indeed, if the airport in question is the
home-base, such a single flight constitutes a separate (trivial)
subproblem, representing an entire forced rotation. Typically, one of
the subproblems will be much larger than the rest, and will be
referred to as the {\em kernel problem}, while the remaining
subproblems will be essentially trivial.

\begin{figure}[t]
\centering
\mbox{  \psfig{figure=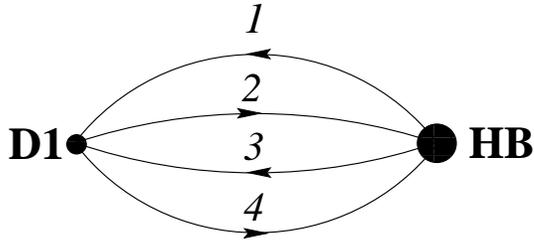,width=7cm}}
\caption{The kernel of the toy problem. The flights have been
relabeled (see table \protect{\ref{effdata}}).}
\label{kernel}
\end{figure}

In this way, our toy problem decomposes into two independent
subproblems, one containing the single composite flight 9-10-11, the
other containing the flights 1-2-3, 4-5, 6-7, and 8. The latter
defines the kernel problem for our toy example. Relabeling the
composite flights gives the structure shown in fig. \ref{kernel}.
In the formalism below, we allow for the possibility that the problem
to be solved has been reduced as described above, which means that
flights may be composite. In what follows we
limit ourselves to the kernel problem.

\section{Optimization with Feedback Neural Networks}

In this section we give a mini-review of how to map resource
allocation problems onto feedback neural networks and the MF
methodology for finding good solutions to such systems. Much of the
formalism here originates from spin models in physics. Hence we will
initially denote the basic degrees of freedom ``spins''.  After
discussing the MF approximation the term ``neuron'' will be used.  We
start out with a binary (Ising) system and then proceed to a
multi-valued (Potts) system. The latter is the most relevant for the
crew scheduling problem.

\subsection{The Ising System}

The Ising system is defined by the energy function
\beq
\label{e_ising}
	E = -\frac{1}{2} \sum_{ij} w_{ij} s_i s_j ,
\eeq
where the binary spins $\si$=$\pm 1$ represent local magnetic
properties of some material, and $\omega_{ij}$ how these spins couple
to each other. Minimizing $E$ in eq. (\ref{e_ising}) yields the spin
configuration of the system.

Feedback networks for resource allocation problems with binary
variables have a similar form. One such example is the graph bisection
problem, where $s_i$ encodes to what partition node $i$ is assigned
and $w_{ij}$=0,1 defines the problem in terms of whether $i$ and $j$
are connected or not. To enforce equal partition, $\sum s_i=0$,
eq. (\ref{e_ising}) needs to be augmented with a soft penalty term.
One gets:
\beq
\label{e2}
	E = -\frac{1}{2} \sum_{ij} w_{ij} s_i s_j +
	\frac{\alpha}{2} \( \sum_i s_i \)^2 ,
\eeq
equivalent to making the replacement $w_{ij} \rightarrow w_{ij} -
\alpha$. The imbalance parameter $\alpha$ sets the relative strength
between the cutsize and the balancing term.  The next step is to find
an efficient procedure for minimizing the energy in
eqs. (\ref{e_ising},\ref{e2}) aiming for the global minimum.

\subsection{Ising Mean Field Equations}

If one attempts to minimize the energy of eq. (\ref{e2}) according to
a local optimization rule, the system will very likely end up in a
local minimum close to the starting point, which is not desired. A
better method is to use a stochastic algorithm that allows for uphill
moves.  One such method is {\it simulated annealing} ({\bf SA})
\cite{kirk}, in which a sequence of configurations is generated,
emulating the Boltzmann distribution
\beq
\label{bz}
	P[s]= \frac{1}{Z} e^{-E[s]/T} ,
\eeq
with neighbourhood search methods. In eq. (\ref{bz}),
$Z$ is the {\it partition function},
\beq
\label{Z1}
	Z = \sum_{[s]} e^{-E[s]/T} ,
\eeq
needed for normalization, and the width or {\em temperature} $T$
represents the noise level of the system. For $T \to 0$ the
Boltzmann distribution collapses into a delta function around the
configuration minimizing $E$. By generating configurations at
successively lower $T$ (annealing) these are less likely to get stuck
in local minima than if $T=0$ from the start. Needless to say, such a
procedure can be very CPU consuming.

The mean field (MF) approach aims at approximating the stochastic SA
method with a deterministic process. This can be derived in two
steps. First $Z$ in eq. (\ref{Z1}) is rewritten in terms of an
integral over new continuous variables $u_i$ and $v_i$.  Then $Z$ is
approximated by the maximum value of its integrand.

To this end, introduce a new set of real-valued variables $v_i$, one
for each spin, and set them equal to the spins with a Dirac
delta-function. Then we can express the energy in terms of the new
variables, and $Z$ takes the form
\beq
\label{Z2}
	Z = \sum_{[s]}\int d[v] e^{-E[v]/T} \prod_i \delta(s_i - v_i)
	= \sum_{[s]} \int d[v] \int d[u] e^{-E[v]/T} \prod_i e^{u_i(s_i - v_i)} ,
\eeq
where the delta-functions have been rewritten by introducing a new
set of variables $u_i$. Then carry out the original sum over $[s]$
and write the product as a sum in the exponent:
\beq
\label{Z4}
	Z \propto \int d[v] \int d[u] e^{-E[v]/T - \sum_i u_i v_i +
	\sum_i \log \cosh u_i} .
\eeq
The original partition function is now rewritten entirely in terms of
the new variables $[u,v]$, with an effective energy in the exponent.
So far no approximation has been made. We next approximate $Z$ in
eq. (\ref{Z2}) by the extremal value of the integrand obtained for
\beq
\label{mft3}
	v_i  =  \tanh \( u_i \) =
	\tanh \( -\frac{\partial E[v]}{\partial v_i} / T \) \; .
\eeq
The mean field variables (or {\it neurons}) $v_i$ can be seen as
approximations to the thermal averages $\langle s_i \rangle_T$ of the
original binary spins.
The MF equations
(eq. (\ref{mft3})) are solved iteratively, either synchronously or
asynchronously, under annealing in $T$.  This defines a feedback ANN.

The dynamics of such an ANN typically exhibits a behaviour with two
phases: When the temperature $T$ is high, the sigmoid function,
$\tanh(\cdot/T)$ in eq. (\ref{mft3}), becomes very smooth, and the
system relaxes into a trivial fixed point, $v_i^{(0)}=0$. As the
temperature is lowered a phase transition (bifurcation) occurs at
$T=T_c$, where $v_i^{(0)}$ becomes unstable, and as $T \rightarrow 0$,
fixed points $v_i^{(*)} = \pm 1$ emerge representing a specific
decision made as to the solution to the optimization problems in
question.
The position of $T_c$ depends upon $w_{ij}$ and can be estimated by
linearizing the sigmoid around $v_i^{(0)}$, i.e. linearizing
eq. (\ref{mft3}).  Based on such an analysis, one can devise a
reliable, parallelizable ``black box'' algorithm for solving problems
of this kind.

Very good numerical results have been obtained for the graph bisection
problem (see ref. \cite{pet2} for references) for a wide range of
problem sizes.
The solutions are comparable in quality to those of the SA method, but
the CPU time consumption is lower than any other known method of
comparable performance. The approach of course becomes even more
competitive with respect to time consumption if the intrinsic
parallelism is exploited on dedicated hardware.

The MF approach differs fundamentally from many other heuristics, in 
that the evolution of the solutions starts outside the proper state 
space, and then gradually approaches the hypercube corners in solution 
space. This feature indicates a relation to interior point methods
\cite{kar}. Indeed, as was pointed out in ref. \cite{yuille},
if the effective (or free) energy is convex, a variant of MF 
annealing can be obtained, which is equivalent to the interior point 
method \cite{kar}.
%

\subsection{The Potts System}

For graph bisection and many other optimization problems, an encoding
in terms of binary elementary variables is natural.  However, there
are many problems where this is not the case. In many cases it is more
natural to replace the two-state Ising spins by multi-valued Potts
spins, which have $K$ possible values (states). For our purposes, the
best representation of a Potts spin is in terms of a vector in the
Euclidean space ${\cal E}_K$. Thus, denoting a spin variable by
$\sss=(s_1,s_2,\ldots,s_K)$, the $j$:th possible state is given by the
$j$:th principal unit vector, defined by $s_j=1$, $s_k=0$ for $k\neq
j$.  These vectors point to the corners of a regular $K$-simplex. They
are each normalized and fulfill the condition
\beq
\label{potts}
	\sum_k s_k = 1,
\eeq
and they are mutually orthogonal,

\subsection{Potts Mean Field Equations}

The MF equations for a system of K-state Potts spins
$\si=(s_{i1},s_{i2},\ldots,s_{iK})$ with an
energy $E(\sss)$ are derived following the same path as in the Ising
case -- rewrite the partition function as an integral over
$\vi$ and $\ui$ and approximate it with the maximum value of the
integrand. One obtains
\begin{eqnarray}
\label{mfp1}
	u_{ij} & = & - \frac{\partial E(\vv)}{\partial v_{ij}} / T , \\
\label{mfp2}
	v_{ij} & = & \frac{e^{u_{ij}}}{\sum_k e^{u_{ik}}},
\end{eqnarray}
from which it follows that the MF {\em Potts neurons} $\vi$, which
approximate the thermal average of $\si$, satisfy
\beq
\label{potts_mf}
	v_{ij} > 0 \, , \; \sum_j v_{ij} = 1.
\eeq
One can think of the neuron component $v_{ij}$ as the {\em
probability} for the $i$:th Potts spin to be in state $j$. For $K=2$
one recovers the formalism of the Ising case in a slightly disguised
form.

Supplying the Potts neurons with a dynamics based on iterating
eqs. (\ref{mfp1},\ref{mfp2}), yields a Potts ANN.  Again one can
typically analyze the linearized dynamics in order to estimate the
critical temperature $T_c$. We refer the reader to ref. \cite{pet2}
for details.

It is often advantageous to replace the derivative in eq. (\ref{mfp1})
with the corresponding difference,
\beq
\label{mfp1_d}
	u_{ij} = - \left ( E|_{v_{ij}=1} - E|_{v_{ij}=0} \right ) / T,
\eeq
which will be used in the airline crew problem below.

\section{Potts Neural Approach to the Crew Scheduling Problem}

\subsection{Encoding}

We are now ready to encode the airline crew problem in terms of Potts
spins. A naive way to do this would be to mimic what was done in the
teachers-and-classes problem in refs. \cite{gis1,gis2}, where each
event (lecture) was mapped onto a resource unit (lecture-room +
time-slot). This would require a Potts spin for each flight to handle
the mapping onto crews.

Since the problem consists in linking together sequences of
(composite) flights into rotations, it appears more natural to choose
an encoding where each flight $i$ is mapped, via a Potts spin, onto
the flight $j$ to follow it in the rotation:
\[
	\sij =
	\left\{ \begin{array}{ll}
		$1$ & \mbox{if flight $i$ precedes flight $j$ in a rotation}, \\
		$0$ & \mbox{otherwise},
	\end{array} \right.
\]
where it is understood that $j$ be restricted to depart from the
(effective) airport where $i$ arrives.  In order to ensure that proper
rotations are formed, each flight has to be mapped onto precisely one
other flight.  This restriction is inherent in the Potts spin
formulation, which is defined to have precisely one component ``on'',
as is evident from eq.  (\ref{potts}).

To start or terminate a rotation, we introduce dummy flights $a$ and
$b$ of zero duration and intrinsic leg count, available only at the
home-base, representing the start/end of a rotation -- at the home-base, $a$ is
formally mapped onto every departure, and every arrival is mapped onto
$b$.

We illustrate the Potts encoding by one particular solution to the toy
kernel problem of fig. \ref{kernel}, where flight 1 is connected to
flight 2, and flight 3 to flight 4. In eq. (\ref{s_toy}) the
``border'' entries of $\sss$ corresponding to the dummy flights $a$
($i,j=0$) and $b$ ($i,j=N+1$) are marked in bold face.
\beq
	\sss =
	\( \begin{array}{cccccc}
		{\bf 0}& {\bf 1}& {\bf 0}& {\bf 1}& {\bf 0}& {\bf 0} \\
		{\bf 0}&      0 &      1 &      0 &      0 & {\bf 0} \\
		{\bf 0}&      0 &      0 &      0 &      0 & {\bf 1} \\
		{\bf 0}&      0 &      0 &      0 &      1 & {\bf 0} \\
		{\bf 0}&      0 &      0 &      0 &      0 & {\bf 1} \\
		{\bf 0}& {\bf 0}& {\bf 0}& {\bf 0}& {\bf 0}& {\bf 0}
	\end{array} \),
\label{s_toy}
\eeq
Global topological properties, leg-counts and durations of rotations,
etc., cannot be described in a simple way by polynomial functions of
the spins. Instead, they are conveniently handled by means of a {\em
propagator} matrix ${\bf P}$, defined in terms of the Potts spin
matrix ${\bf s}$ by
\beq
\label{prop}
	\Pij = \( ( {\bf 1}-\sss )^{-1} \)_{ij} = \delta
		_{ij} + \sij + \sum_k s_{ik} s_{kj} + \sum_{kl}
		s_{ik}s_{kl}s_{lj} + \sum_{klm} s_{ik}s_{kl}
		s_{lm}s_{mj} + \ldots
\eeq
\begin{figure}[htb]
\centering
\mbox{\psfig{figure=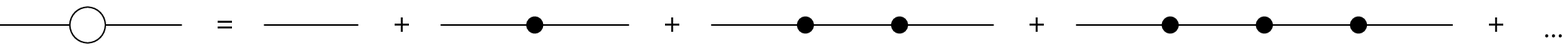,width=14.8cm,height=.4cm}}
\vspace{0.2in}
\caption{Expansion of the propagator $\Pij$ ($\bigcirc$) in terms of
 	$\sij$. A line represents a flight, and ($\bullet$) a
 	landing.}
\label{fig_prop}
\end{figure}
A pictorial expansion of the propagator is shown in
fig. \ref{fig_prop}. The interpretation is obvious: $\Pij$ counts
the number of connecting paths from flight $i$ to $j$. The P-matrix
corresponding to the toy problem solution of eq. (\ref{s_toy}) is
given by
\beq
	\PP =
	\left(
	\begin{array}{cccccc}
		{\bf 1}& {\bf 1}& {\bf 1}& {\bf 1}& {\bf 1}& {\bf 2} \\
		{\bf 0}&      1 &      1 &      0 &      0 & {\bf 1} \\
		{\bf 0}&      0 &      1 &      0 &      0 & {\bf 1} \\
		{\bf 0}&      0 &      0 &      1 &      1 & {\bf 1} \\
		{\bf 0}&      0 &      0 &      0 &      1 & {\bf 1} \\
		{\bf 0}& {\bf 0}& {\bf 0}& {\bf 0}& {\bf 0}& {\bf 1}
	\end{array}
	\right),
\label{p_toy}
\eeq
where one finds two paths from $a$ to $b$ ($P_{ab} = 2$), as there
should be -- one via 1-2 and one via 3-4.

Similarly, an element of the matrix square of $\PP$,
\beq
\label{P2}
	\(P^2\)_{ij} \equiv \sum_k P_{ik} P_{kj}
                   = \delta_{ij} + 2 s_{ij} + 3 \sum_k s_{ik}s_{kj} + \ldots ,
\eeq
counts the total number of composite legs in the connecting paths
between $i$ and $j$, while the number of proper legs is given by
\beq
\label{PLP}
	\tilde{L}_{ij} \equiv \sum_k P_{ik} L_k P_{kj}
	= \delta_{ij} L_i
	+ s_{ij} \( L_i + L_j \)
	+ \sum_k s_{ik} s_{kj} \( L_i + L_k + L_j \)
	+ \ldots ,
\eeq
where $L_k$ is the intrinsic number of single legs in the composite
flight $k$. For the toy model solution, we get
\beq
	\PP^2 =
	\left(
	\begin{array}{cccccc}
		{\bf 1}&{\bf 2}&{\bf 3}&{\bf 2}&{\bf 3}&{\bf 8} \\
		{\bf 0}&     1 &     2 &     0 &     0 &{\bf 3} \\
		{\bf 0}&     0 &     1 &     0 &     0 &{\bf 2} \\
		{\bf 0}&     0 &     0 &     1 &     2 &{\bf 3} \\
		{\bf 0}&     0 &     0 &     0 &     1 &{\bf 2} \\
		{\bf 0}&{\bf 0}&{\bf 0}&{\bf 0}&{\bf 0}&{\bf 1}
	\end{array}
	\right)
\eeq
and
\beq
	\tilde{\LL} =
	\left(
	\begin{array}{cccccc}
		{\bf 0}&{\bf 3}&{\bf 5}&{\bf 2}&{\bf 3}&{\bf 8} \\
		{\bf 0}&     3 &     5 &     0 &     0 &{\bf 5} \\
		{\bf 0}&     0 &     2 &     0 &     0 &{\bf 2} \\
		{\bf 0}&     0 &     0 &     2 &     3 &{\bf 3} \\
		{\bf 0}&     0 &     0 &     0 &     1 &{\bf 1} \\
		{\bf 0}&{\bf 0}&{\bf 0}&{\bf 0}&{\bf 0}&{\bf 0}
	\end{array}
	\right) ,
\eeq
based on the intrinsic leg counts 3, 2, 2, and 1, for the flights 1,
2, 3, and 4, respectively. The path via 1-2 has 5 legs, and the one
via 3-4 has 3 legs, making a total leg count of 8 for all paths from
$a$ to $b$, as can be read off from the upper right corner
($\tilde{L}_{ab} = 8$).

The {\em average leg count} of the connecting paths is then given by
\beq
\label{Lij}
	L_{ij} \equiv \frac{\tilde{L}_{ij}}{\Pij},
\eeq
and for the {\em average duration} (flight + waiting-time) of the
paths from $i$ to $j$ one has
\beq
\label{Tij}
	T_{ij} \equiv \frac{\sum_k P_{ik} t^{\mf{(f)}}_k P_{kj} +
	\sum_{kl} P_{ik} t^{\mf{(w)}}_{kl} s_{kl} P_{lj}}
	{P_{ij}},
\eeq
where $t^{\mf{(f)}}_i$ denotes the duration of the composite flight
$i$, including the embedded waiting-time. The averaging is
accomplished by the division with $P_{ij}$.
In principle, this could lead to undefined expressions in cases with
$P_{ij} = 0$. This will be no problem, since we will only be
interested in cases either with $i=a$, probing the path from $a$ to
$j$, i.e. the part up to $j$ of the rotation containing $j$, or $j=b$,
probing the path from $i$ to $b$, i.e. the part after and including
$i$ of the rotation to which $i$ belongs.

Furthermore, any improper loops (such as obtained e.g. if two flights
are mapped onto each other) will make ${\bf P}$ singular -- for a
proper set of rotations, $\det {\bf P} = 1$.

\subsection{Mean Field Dynamics}

In the MF formalism the basic dynamical variables are $\vv$ rather
than $\sss$; correspondingly, we will use a probabilistic propagator
$\PP$, defined as the matrix inverse of ${\bf 1} - \vv$, in analogy
with eq. (\ref{prop}), but with $\sss$ replaced by $\vv$.  The
clearcut structure seen in the toy-model matrices in
eqs. (\ref{s_toy},\ref{p_toy}), will only emerge as $T \to 0$.

Rather than finding a suitable energy function in terms of the
matrices $\vv$ and $\PP$, we have chosen a more pragmatic approach by
directly writing down the {\it local fields} $u_{ij}$, bypassing
eq. (\ref{mfp1}). The corresponding mean fields $v_{ij}$ are obtained
from the MF equations (eq. (\ref{mfp2})); they have an obvious
interpretation of probabilities (for flight $i$ to be followed by
$j$).

In the MF equations (eq. (\ref{mfp2})) $v_{ij}$ will be updated for
one flight $i$ at a time, by first zeroing the $i$:th row of $\vv$
(and updating $\PP$ correspondingly), and then computing the relevant
local fields $\uij$ entering eq. (\ref{mfp2}) as
\beq
\label{uij}
   \uij =
    - \frac{c_1}{T} t_{ij}^{\mf{(w)}}
    - \frac{c_2}{T} \sum_k v_{kj}
    - \frac{c_3}{T} \log\(\frac{1}{1 - P_{ji}}\)
    - \frac{c_4}{T} \Psi \( T^{\mf{(ij)}}_{\mf{rot}} - T^{\mf{max}}_{\mf{rot}} \)
    - \frac{c_5}{T} \Psi \( L^{\mf{(ij)}}_{\mf{rot}} - L^{\mf{max}}_{\mf{rot}} \),
\eeq
where $j$ is restricted to be a possible continuation flight to $i$.
It is difficult, and not necessary from the viewpoint of algorithmic
performance, to find energy functions corresponding to the fourth and
fifth terms in eq. (\ref{uij}). In contrast, the first and second
terms are straightforward in this respect, and the third term
originates from an energy term $\sim \log\mbox{det} \,\PP$.  The five
different terms in eq. (\ref{uij}) serve the following purposes:
\begin{enumerate}
\item Cost term: The local waiting-time $t_{ij}^{\mf{(w)}}$ between
  flight $i$ and $j$.
\item Penalizes solutions where two flights point to the same next
  flight.
\item Suppresses improper loops. $P_{ji} \to 1$ if a path $j \to i$
  exists, i.e. if a loop is formed if $i$ connects with $j$. The
  penalty approaches $\infty$ when $P_{ji} \to 1$.
\item Prohibits violation of the bound $T^{\mf{max}}_{\mf{rot}}$ on
  total rotation time, where $T^{\mf{(ij)}}_{\mf{rot}}$ stands for the
  duration of the resulting rotation if $i$ where to be mapped onto
  $j$.
\item Prohibits violation of the bound $L^{\mf{max}}_{\mf{rot}}$ on
  total number of legs, where $L^{\mf{(ij)}}_{\mf{rot}}$ is the
  resulting number of legs in the rotation if $i$ where to be mapped
  onto $j$.
\end{enumerate}

In eq. (\ref{uij}), the rotation time $T^{\mf{(ij)}}_{\mf{rot}}$ and
the leg count $L^{\mf{(ij)}}_{\mf{rot}}$ are given as
\begin{eqnarray}
\label{T}
	T^{\mf{(ij)}}_{\mf{rot}} & = & T_{ai} + t^{\mf{(w)}}_{ij} + T_{jb},
\\ \label{L}
	L^{\mf{(ij)}}_{\mf{rot}} & = & L_{ai} + L_{jb},
\end{eqnarray}
in terms of eqs. (\ref{Tij},\ref{Lij}).

The penalty function $\Psi$, used to enforce the inequality
constraints \cite{ohl}, is defined by $\Psi(x) = x \Theta(x)$ where
$\Theta$ is the Heaviside step function. It turns out, as will
be discussed below, that the performance of the algorithm is
fairly insensitive to the choice of the relative strengths
$c_i$ occurring in eq. (\ref{uij}).

After an initial computation of the propagator ${\bf P}$ from scratch, it
is subsequently updated according to the Sherman-Morrison algorithm
for incremental matrix inversion \cite{numrec}. An update of the $i$:th
row of ${\bf v}$, $v_{ij} \rightarrow v_{ij} + \delta_j$, generates
precisely the following change in the propagator ${\bf P}$:
\beq
\label{sher1}
	P_{kl}  \rightarrow  P_{kl} + \frac{P_{ki} \, z_{l}}{1-z_{i}},
\eeq
with
\beq
\label{sher2}
	 z_{l}  =  \sum_{j} \delta_{j} \, P_{jl}.
\eeq
Inverting the matrix from scratch would take $O(N^3)$ operations,
while the (exact) scheme devised above only requires $O(N^2)$ per row.

In principle, a proper value for an initial temperature can be
estimated from linearizing the dynamics of the MF equations. The
neurons are initialized close to the trivial fixed point.  A common
annealing schedule for the updating, based on iterating the MF
eqs. (\ref{mfp2},\ref{sher1}), is to decrease $T$ by a fixed factor
per iteration.

As the temperature goes to zero, a solution crystallizes in a
winner-takes-all dynamics: for each flight $i$, the largest
$u_{ij}$ determines the continuation flight $j$ to be chosen.

Implementation details of the algorithm can be found in Appendix B.

\subsubsection{Parallelizing the Algorithm}

One obstacle, if one wants to parallelize the algorithm, is that the
scheme above, eqs. (\ref{sher1},\ref{sher2}), for updating $\PP$ is
non-local, in that all matrix elements of $\PP$ are updated due to a
change in a single neuron $\vv_i$.
An alternative method using only local information on
$\vv$ and $\PP$ is to update row $i$ of ${\bf P}$ according to
\beq
\label{psoft}
	P_{im} \rightarrow \delta_{im} + \sum_{j} \vij P_{jm},
\eeq
in connection with updating the corresponding row of $\vv$. If each
flight keeps track of its own row of $\PP$, all information needed can
be obtained from the possible continuation flights $j$ (``neighbours'')
to flight $i$. This scheme gives convergence towards the exact
inverse; a similar method has been successfully used in the context of
communication routing \cite{hakk}.

The handling of the global time and leg constraints of a rotation
could be tackled in a similar manner, with each flight keeping track
of the time and number of legs used both from $a$ to itself and
from itself to $b$, where $a$ and $b$ are the dummy flights starting
and terminating a rotation.
The information needed to calculate $T^{\mf{(ij)}}_{\mf{rot}}$ and
$L^{\mf{(ij)}}_{\mf{rot}}$ then is local to $i$ and its ``neighbours''
$j$.
\section{Test Problems}

In choosing test problems our aim has been to maintain a reasonable
degree of realism, while avoiding unnecessary complication and at the
same time not limiting ourselves to a few real-world problems, where
one can always tune parameters and procedures to get a good
performance. In order to accomplish this we have analyzed two typical
real-world {\it template} problems obtained from a major airline: one
consisting of long distance (LD), the other of short/medium distance
(SMD) flights.
\begin{figure}
\centering
\mbox{
\psfig{figure=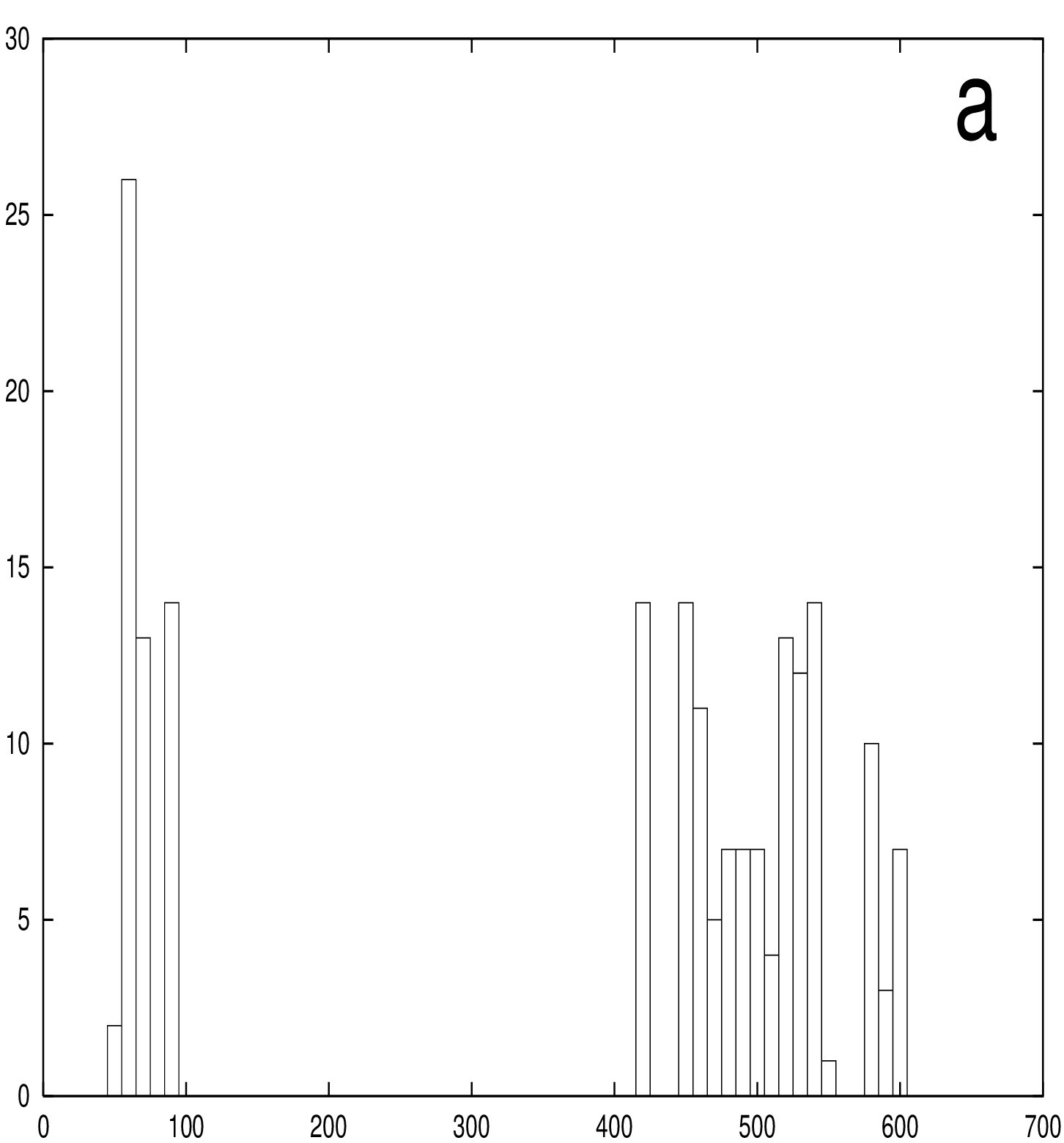,width=7.4cm,height=5cm}
\psfig{figure=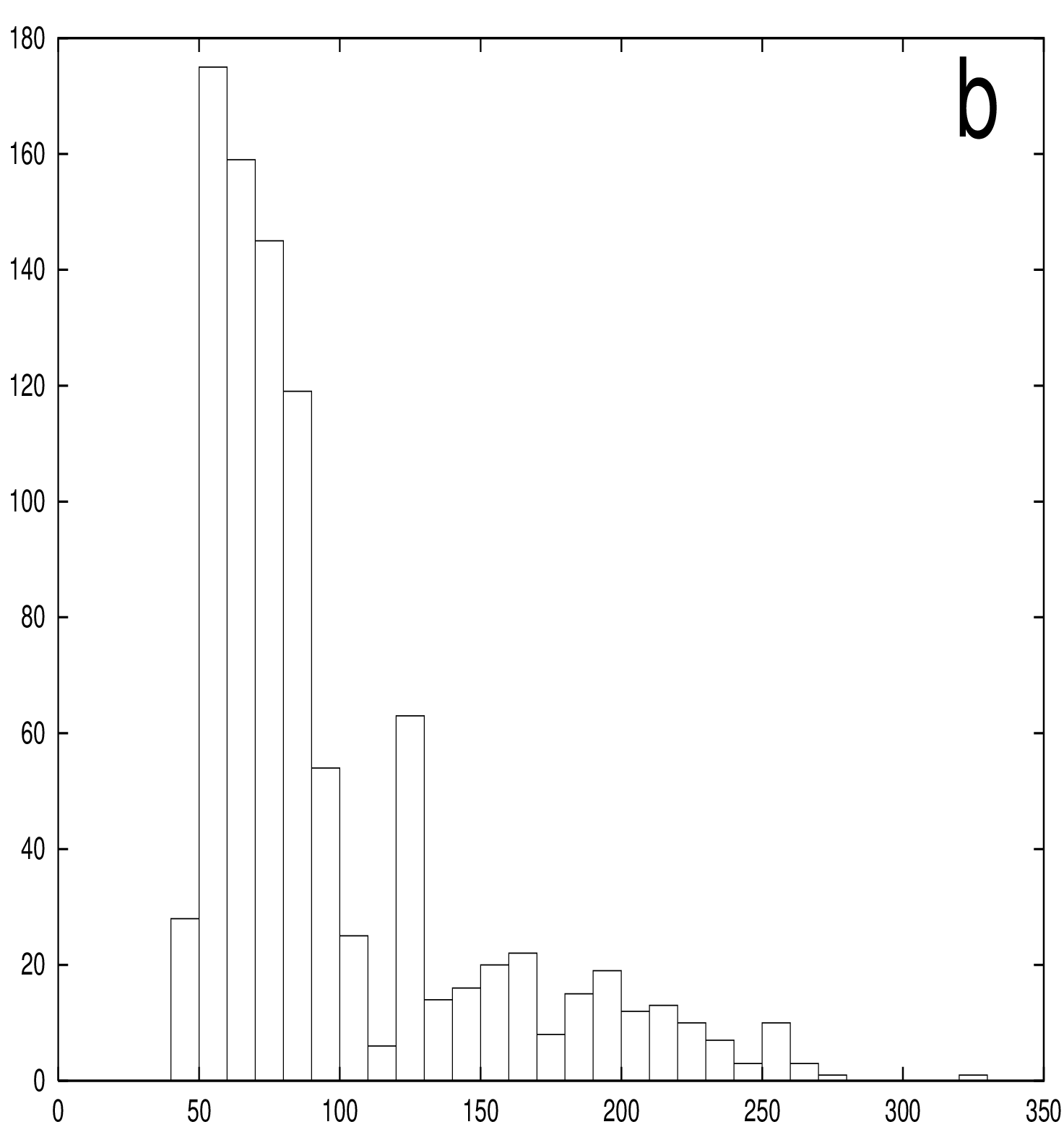,width=7.4cm,height=5cm}
}
\caption{Flight time distributions in minutes for {\bf (a)} LD and {\bf (b)}
	SMD template problems.}
\label{fig_lh}
\end{figure}
As can be seen from fig. \ref{fig_lh}, LD flight time distributions
are centered around long times, with a small hump for shorter times
representing local continuations of long flights. The SMD flight times
have a more compact distribution.

For each template we have made a distinct problem generator producing
random problems resembling the template. For algorithm details see
Appendix C.

Due to the excessive time consumption
of the available exact methods, the performance of the Potts approach
cannot be tested against these -- except for in this context
quite small problems, for which the Potts solution quality
matches that of an exact algorithm. For artificial problems of more
realistic size we circumvent this obstacle in the following way: since
problems are generated by producing a legal set of rotations, we add
in the generator a final check that the implied solution yields
$T^{\mf{unr}}_{\mf{wait}}$; if not, a new problem is
generated. Theoretically, this might introduce a bias in the problem
ensemble; empirically, however, no problems have had to be
redone. Also the two template problems turn out to be solvable at
$T^{\mf{unr}}_{\mf{wait}}$.
%
%

Each problem then is reduced as described above (using a negligible
amount of computer time), and the kernel problem is
stored as a list of flights, with all traces of the generating
rotations removed.

\section{Results}

We have tested the performance of the Potts MF approach for both LD
and SMD kernel problems of varying sizes.

\begin{table}[ht]
\begin{center}
\begin{tabular}{|c|c|c|c|c|c|}
\hline
$c_1$ & $c_2$ & $c_3$ & $c_4$ & $c_5$ \\
\hline
$\mbox{period}^{-1}$ & $1$ & $1$ & $\langle T^{\mf{rot}}\rangle^{-1}$ & $\langle L^{\mf{rot}}\rangle^{-1}$ \\
\hline
\end{tabular}
\end{center}
\caption{The coefficients used in
	eq. (\protect\ref{uij}). $\langle T^{\mf{rot}}\rangle$ is the average
	duration per rotation (based on $T^{\mf{unr}}_{\mf{wait}}$),
	and $\langle L^{\mf{rot}}\rangle$ the average leg count, both of which can
	be computed beforehand.}
\label{c}
\end{table}
The values used for the coefficients $c_i$ in eq. (\ref{uij}) are
displayed in table \ref{c}.
One should stress that these parameter settings have been used for the
entire range of problem sizes probed. For the LD problems the bounds
on a rotation are chosen as
\beqa
	T^{\mf{max}}_{\mf{rot}} & = & 10000,\\
	L^{\mf{max}}_{\mf{rot}} & = & 15,
\eeqa
and to
\beqa
	T^{\mf{max}}_{\mf{rot}} & = & 6000,\\
	L^{\mf{max}}_{\mf{rot}} & = & 25,
\eeqa
for the SMD problems.

\begin{figure} [htb]
\centering
\mbox{\psfig{figure=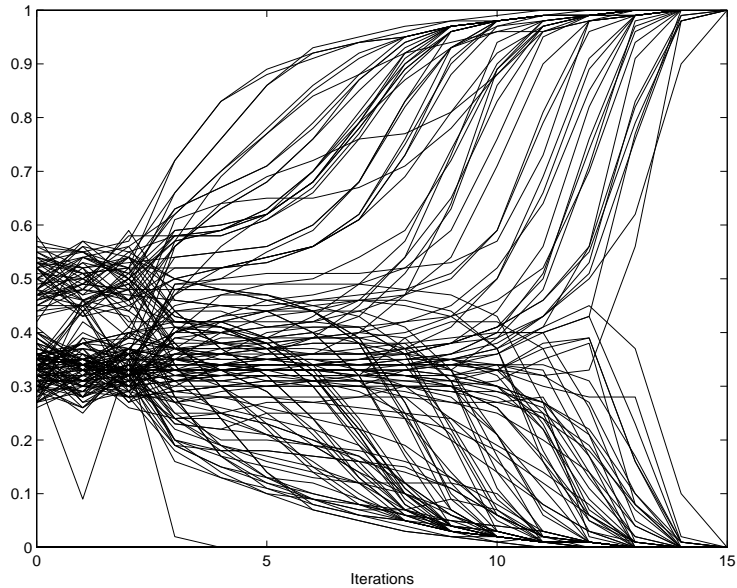,width=10cm}}
\caption{Evolution of neuron components $v_{ij}$ as the temperature
is lowered for the template LD problem.}
\label{evol}
\end{figure}

\begin{figure} [htb]
\centering \mbox{\psfig{figure=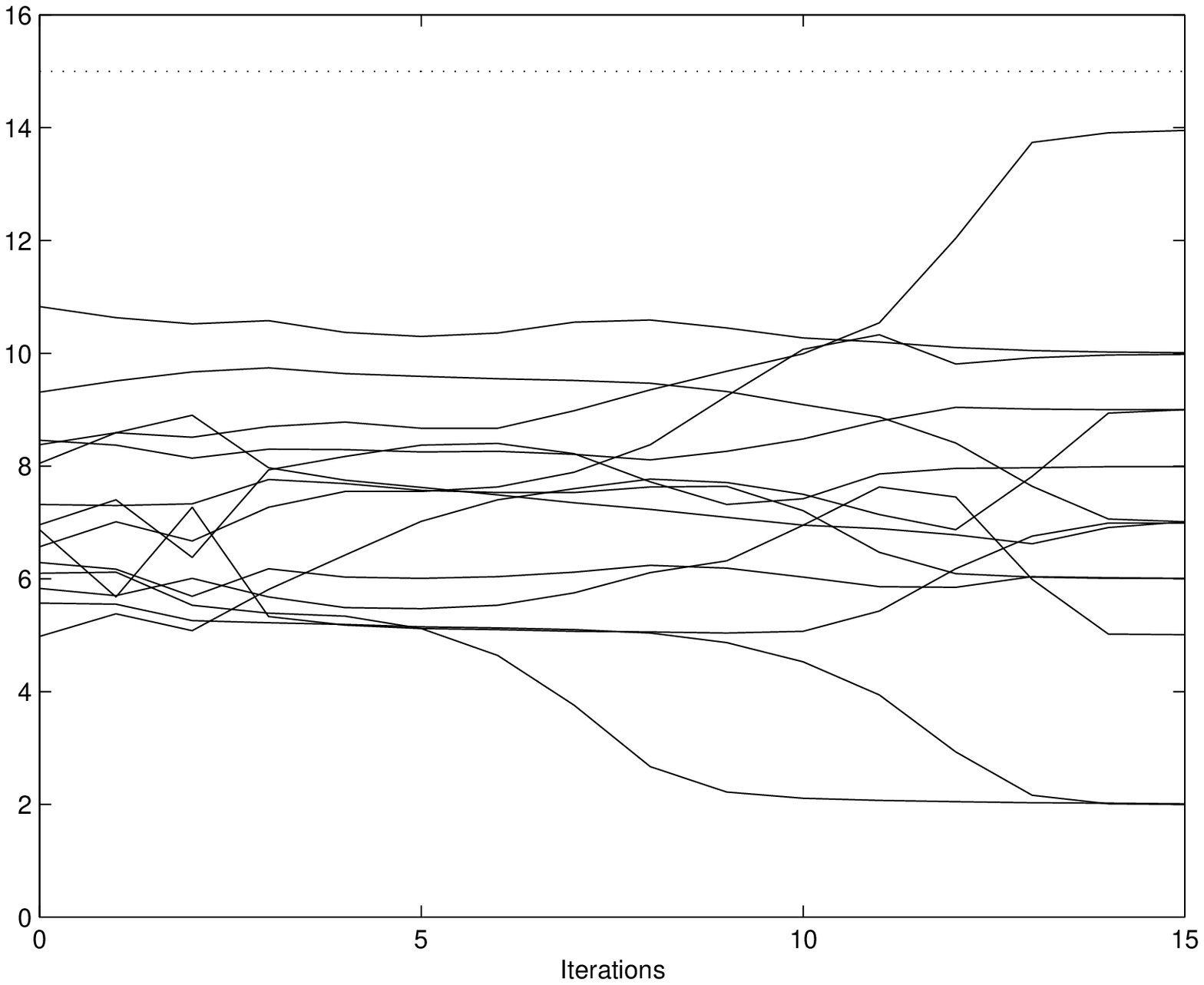,width=7.5cm}
\psfig{figure=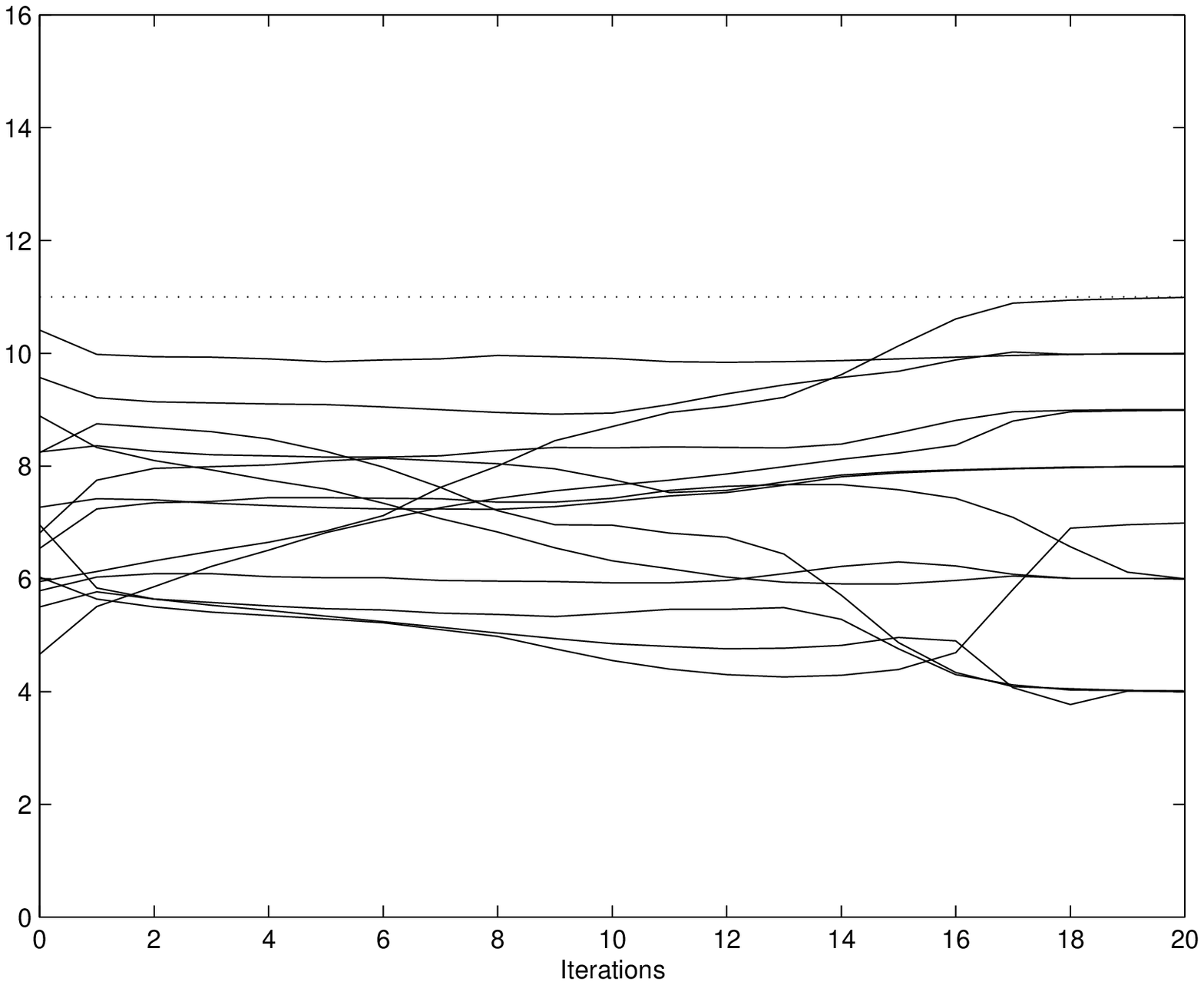,width=7.5cm}}
\caption{Evolution of $L_{ib}$ for all departure flights from the home
base, $i$, for the template LD problem. The dotted lines denote
$L^{\mf{max}}_{\mf{rot}}$.}
\label{legs}
\end{figure}

A typical evolution of the individual neuron components $v_{ij}$ is
shown in fig. \ref{evol}. In fig. \ref{legs} the evolution of the
number of legs of all the rotations (defined by $L_{ib}$ where $i$ is
a departure flight from HB) for two different values of the bound
$L^{\mf{max}}_{\mf{rot}}$.

When evaluating a solution obtained with the Potts approach, a check
is done as to whether it is legal (if not, a simple post-processor
tries to restores legality -- this is only occasionally needed), then the
solution quality is probed by measuring the excess waiting-time $R$,
\beq
\label{perf}
	R = \frac{T^{\mf{wait}}_{\mf{Potts}}-T^{\mf{unr}}_{\mf{wait}}}
	{\mbox{period}},
\eeq
which is a non-negative integer for a legal solution.

For a given problem size, as given by the desired number of airports
$N_a$ and flights $N_f$, a set of 10 distinct problems is generated.
Each problem is subsequently reduced, and the Potts algorithm is
applied to the resulting kernel problem.  The solutions are evaluated,
and the average $R$ for the set is computed.  The results for a set of
problem sizes ranging from $N_f\simeq 75$ to $1000$ are shown in
tables \ref{fig_LD} and \ref{fig_SMD}; for the two template problems
see table \ref{fig_ori}.
%
%
%
%
\begin{table}[t]
\begin{center}
\begin{tabular}{|c|c|c|c|c|c|}
\hline
$N _f$ & $N_a$ & $< N^{\eff}_f>$ & $< N^{\eff}_a>$ & $<$ R $>$ & $< $ CPU time $ >$\\
\hline
 75 &  5 &  23 &  8 & 0.0 & 0.0 sec  \\ 
100 &  5 &  44 & 13 & 0.0 & 0.2 sec  \\
150 & 10 &  46 & 14 & 0.0 & 0.1 sec  \\
200 & 10 &  84 & 24 & 0.0 & 0.7 sec  \\
225 & 15 &  74 & 22 & 0.0 & 0.4 sec  \\
300 & 15 & 132 & 38 & 0.0 & 1.5 sec  \\
\hline
\end{tabular}
\end{center}
\caption{Average performance of the Potts algorithm for {\bf LD}
	problems. The superscript ``eff'' refers to the kernel
	problem, subscripts ``$f$'' and ``$a$'' refer to respectively
	flight and airport. The averages are taken with 10 different
	problems for each $N_f$. $R$ is the excess waiting-time, and
	the CPU time refers to DEC Alpha 2000.}
\label{fig_LD}
\end{table}
\begin{table}[ht]
\begin{center}
\begin{tabular}{|c|c|c|c|c|c|}
\hline
$N _f$ & $N_a$ & $< N^{\eff}_f>$ & $< N^{\eff}_a>$ & $<$ R $>$ & $< $ CPU time $ >$\\
\hline
600 & 40  & 280 &  64 & 0.0 &  19 sec  \\
675 & 45  & 327 &  72 & 0.0 &  35 sec  \\
700 & 35  & 370 &  83 & 0.0 &  56 sec  \\
750 & 50  & 414 &  87 & 0.0 &  90 sec  \\
800 & 40  & 441 &  91 & 0.0 & 164 sec  \\
900 & 45  & 535 & 101 & 0.0 & 390 sec  \\
1000& 50  & 614 & 109 & 0.0 & 656 sec  \\
\hline
\end{tabular}
\end{center}
\caption{Average performance of the Potts algorithm for {\bf SMD}
	problems. The averages are taken with 10 different problems
	for each size. Same notation as in table
	\protect\ref{fig_LD}.}
\label{fig_SMD}
\end{table}
\begin{table}[t]
\begin{center}
\begin{tabular}{|c|c|c|c|c|c|c|}
\hline
$N _f$ & $N_a$ & $< N^{\eff}_f>$ & $< N^{\eff}_a>$ & $<$ R $>$ & $< $ CPU time $ >$ & type\\
\hline
189 & 15 &  71 & 24 & 0.0 & 0.6 sec &  LD \\
948 & 64 & 383 & 98 & 0.0 & 184 sec & SMD \\
\hline
\end{tabular}
\end{center}
\caption{Average performance of the Potts algorithm for 10 runs on the
	two template problems. Same notation as in table
	\protect\ref{fig_LD}.}
\label{fig_ori}
\end{table}

The results are quite impressive -- the Potts algorithm has solved all
problems, and with a very modest CPU time consumption,
%
%
of which the major part is used in updating the $P$ matrix, using the
fast method of eqs. (\ref{sher1}, \ref{sher2}). The iteration time scales
like $(N^{\eff}_f)^3 \propto N_f^3$ with a small prefactor. This
should be multiplied by the number of iterations needed -- empirically
between 20 and 40, independently of problem size\footnote{The minor
apparent deviation from the expected scaling in tables
\protect\ref{fig_LD}, \protect\ref{fig_SMD} and \protect\ref{fig_ori}
are due to an anomalous scaling of the Digital DXML library routines
employed; the number of elementary operations does scale like
$N_f^3$.}.

\section{Summary}

We have developed a mean field Potts approach for finding good
solutions to airline crew scheduling problems resembling
real-world situations.

A key feature is the handling of global entities, sensitive to the
dynamically changing ``fuzzy'' topology, by means of a propagator
formalism.
This is a novel ingredient in ANN-based approaches to resource
allocation problems.
Another important ingredient is the problem size reduction achieved
by airport fragmentation and flight clustering, narrowing down the
solution space by removing much of the sub-optimal part.
This is done by exploiting the local properties of the corresponding
unrestricted problems \cite{loc}.

High quality solutions are consistently found throughout a range of
problem sizes without having to fine-tune the parameters, with a time
consumption scaling as the cube of the problem size. The performance
of the Potts algorithm is probed by comparing to the unrestricted
optimal solutions.

At first sight, the Potts algorithm appears difficult to implement in
a parallel way with its global quantities. A concurrent
implementations can be facilitated, however, by localizing all
information.

The basic approach presented here is easy to adapt to other
applications, in particular in communication routing \cite{hakk}.

\subsection*{Acknowledgements:}

We thank Richard Blankenbecler for valuable suggestions on the
manuscript.  This work was in part supported by the Swedish Natural
Science Research Council and the Swedish Board for Industrial and
Technical Development.

\newpage

%
\appendix
\renewcommand{\thesection}{Appendix \Alph{section}.
\setcounter{equation}{0}}
\setcounter{table}{0}
\renewcommand{\theequation}{\Alph{section}\arabic{equation}}
\renewcommand{\thetable}{\Alph{section}\arabic{table}}
\newcommand{\app}[1]{\newpage\section{#1}}

\app{A Toy Example}

In this Appendix we define and analyze a small toy example with five
airports, that is used throughout this paper to illustrate the various
steps. When exploring the performance of the algorithm, much larger
problems are of course involved (see Section 7).

The toy problem is specified
in table \ref{fulldata}, where a period of 10080 is assumed.
\begin{table}[ht]
\begin{center}
\begin{tabular}{|c|c|c|c|c|c|}
\hline
Flight & Dep. airport & Arr. airport & Dep. time & Arr. time \\
\hline
 1  & 	   HB &       B  &     0    &   500  \\
 2  &       B &       C  &     1000 &   1300 \\
 3  &       C &       D  &     1500 &   1850 \\
 4  &       D &       E  &     4300 &   4870 \\
 5  &       E &      HB  &     5100 &   5500 \\
 6  & 	   HB &       B  &     1500 &   2000 \\
 7  &       B &       D  &     2200 &   2800 \\
 8  &       D &      HB  &     3500 &   4100 \\
 9  & 	   HB &       B  &     6000 &   6500 \\
 10 &       B &       D  &     7000 &   7500 \\
 11 &       D &      HB  &     8000 &   8250 \\
\hline
\end{tabular}
\end{center}
\caption{Toy problem specification.}
\label{fulldata}
\end{table}
An illustration is shown in fig. \ref{fullrot}.

Fragmentation and clustering with flight duration times and the number
of legs added together gives the structures shown in
fig. \ref{effrot}b and table \ref{effdata}. The corresponding kernel
problem is shown in fig. \ref{kernel} with the effective flights
relabeled according to table \ref{effdata}.
\begin{table}[htb]
\begin{center}
\begin{tabular}{|c|c|c|c|c|c|c|}
\hline
Comp. flight & Leg-count & Legs & Dep. airport & Arr. airport & Dep. time & Arr. time \\
\hline
 1 & 3 &   1-2-3 &    HB &    D1  & 0        & 1850 \\
 2 & 2 &     4-5 &    D1 &    HB  & 4300     & 5500 \\
 3 & 2 &     6-7 &    HB &    D1  & 1500     & 2800 \\
 4 & 1 &       8 &    D1 &    HB  & 3500     & 4100 \\
 5 & 3 & 9-10-11 &    HB &    HB  & 6000     & 8250 \\
\hline
\end{tabular}
\end{center}
\caption{Toy model description after fragmentation and clustering. The
first column gives the composite flight label.}
\label{effdata}
\end{table}
The total flight-time is 5070, and without restrictions, there are two
solutions with minimum waiting-time, 5280.
One consists in the rotations 1-2, 3-4, and 5 in terms of composite
flights, i.e. 1-2-3-4-5, 6-7-8, and 9-10-11 in terms of the original
flights. The corresponding rotation times are 5500, 2600 and 2250,
respectively.
The other has the rotations 1-4, 2-3, and 5 in terms of composite
flights, i.e. 1-2-3-8, 6-7-4-5, and 9-10-11 in terms of proper
flights, with the respective rotation times 4100, 4000, and 2250.

\app{The Potts Algorithm}

\newcommand{\AlgBox}[2]{
\begin{minipage}[t]{15cm}
{\large\bf #1}\\[3mm]
\framebox[15cm][l]{
\begin{minipage}{14.5cm}#2
\end{minipage}}\end{minipage}
}

\subsection*{Initialization}

The initial temperature $T_0$ is assigned a tentative value of $1.0$.
If the averaged squared change of the neurons,
\beq
\label{B1}
	(\Delta v)^2 = \frac{1}{\tilde{N}_f} \sum_{ij}(\Delta v_{ij})^2 =
	\frac{1}{\tilde{N}_f} \sum_{ij}(v_{ij}(t+1)-v_{ij}(t))^2,
\eeq
is larger than 0.2 after the first iteration, then the system is
reinitialized with $T_0 \rightarrow 2 T_0$. If, on the other hand, it
is smaller than 0.01 the system is reinitialized with $T_0 \rightarrow
T_0 / 2$. In eq. (\ref{B1}) $\tilde{N}_f$ is the number of flights
minus those departing from HB,

Each neuron ${\bf v}_i$ is initialized by assigning random values to
its components $v_{ij}$ in the interval $0.8/K$ to $1.2/K$, where $K$
is the number of components of the Potts neuron. The neuron is then
normalized by dividing each component by the component sum.

Subsequently, $P_{ij}$, $T_{ij}$ and $L_{ij}$ are initialized
consistently with the neuron values.

The following iteration is repeated, until one of the termination
criteria (see below) is fulfilled:
\\[3mm]
\AlgBox{Iteration}
{
  \begin{itemize}
    \item For each airport (in random order) do:
    \begin{enumerate}
      \item For each arrival flight $i$, do:
      \begin{enumerate}
	\item Update ${\bf v}_i$ (eqs. (\ref{mfp2},\ref{uij}))
	\item Update ${\bf P}$ (eq. (\ref{sher1})
      \end{enumerate}
      \item Correct the neuron matrix by doing the following $N_{\mf{norm}}$ times:
      \begin{enumerate}
	\item Normalize the columns of {\bf v}, corresponding to local departures.
	\item Normalize the rows of {\bf v}, corresponding to local arrivals.
      \end{enumerate}
    \end{enumerate}
    \item Decrease the temperature: $T = kT$.
  \end{itemize}
}
\\[3mm]
We have consistently used $k = 0.9$ and $N_{\mf{norm}} = 2$.

\subsection*{Termination criteria}

The updating process is terminated if
\beq
	1/\tilde{N}_f \sum_{ij} v_{ij}^2 > 0.99 \;\;\mbox{and}\;\;
	\max_{ij}(\Delta v_{ij})^2 < 0.01 \;\;\mbox{and}\;\;
	\min_{ij}v_{ij}^2 > 0.8,
\eeq
or
\beq
	1/\tilde{N}_f \sum_{ij} v_{ij}^2 > 0.8 \;\;\mbox{and}\;\;
	(\Delta v)^2 > 0.000001  \;\;\mbox{and}\;\;
	\max_{ij}(\Delta v_{ij})^2 < 0.01 \;\;\mbox{and}\;\;
	\min_{ij}v_{ij}^2 > 0.8,
\eeq
or if the number of iterations exceeds 100.

\subsection*{Postprocessing}

First, the final state of each neuron $\vi$ is analyzed with respect
to its implied choice, defined by its largest component.  A check is
done as to whether a proper rotation structure results. If this is not
the case (which never happened for the problems studied here), one may
e.g. rerun the algorithm with modified parameters.

Then, each rotation is checked for legality: If the rotation time or
leg count exceeds the respective bound, a simple algorithm is employed to
attempt to restore legality, by swapping flights between rotations.
For the few cases ($\sim$ 5 \%) where such a correction was needed for the
problems studied here, this procedure always sufficed.

\app{The Problem Generator}

We have made two distinct problem generators, tuned to generate
problems statistically resembling the LD or SMD templates.  In both
generators, a random problem with a specified number of airports and
flights is generated as follows.

First, the flight-times between airports are chosen randomly from a
distribution, based on the relevant template problem. Then, a flight
schedule is built up in the form of legal rotations starting and
ending at the home-base.
The waiting-times between consecutively flights are chosen in a
random fashion
in the neighbourhood of $T^{\mf{unr}}_{\mf{wait}}/N_f$ for the
corresponding template problem.

For a specified number of airports $N_a$ and flights $N_f$, the key
steps take the following form.

\AlgBox{Generator steps}
{\begin{itemize}
 \item The airports are assigned distinct probabilities, designed to match the traffic
	distribution for the airports in the relevant template problem.
 \item For each pair of airports, a distance (flight-time) is drawn
	from a predefined distribution.
 \item While the number of generated flights is less than $N_f$:\\
	Start a new rotation from HB, then for each leg do:
		\begin{enumerate}
		\item Choose its destination:
                	\begin{itemize}
                	\item If the number of legs is less than $L_0$,
				draw the destination from a predefined distribution,
                		where HB is chosen with a probability $\cal{P}_{\mf{HB}}$.
                		\footnote{Care is taken that the very last leg goes to HB.}
				\footnote{If more than half of the
                		flights are generated and some airports
                		still are not visited, then if the destination is not HB,
                		change to an unvisited airport. Only one
                		airport per rotation is allowed to be
                		chosen in this way.}
			\item Else force the destination to be HB, and
				begin a new rotation.
                	\end{itemize}
		\item Pick the waiting-time from the predefined distribution.
		\item Set the flight time according to the distance table, with some random deviation.
		\end{enumerate}
 \item If any rotation time exceeds the limit, or if the solution does not end
      up at the unrestricted minimal waiting-time, generate a new problem.
\end{itemize}
}

The probability $\cal{P}_{\mf{HB}}$ for choosing HB as a destination
is for both problem types chosen to 0.25, except for the
first leg, giving on the average 5 legs per rotation.
For LD-problems, the maximum legcount $L_0$ is set to 15, while for
the SMD problems it is set to 25.
%
%

\end{document}